\documentclass{article}
\usepackage{graphics}
\begin{document}

\input{newcmnds.lst}  % Author's own list of newcommands

\bc 
{\LARGE\bf  Time, Distance, Velocity, Redshift: a personal 
guided tour}

\vs\vs

{\Large T. Kiang}\\

\vs

{\large\it Dunsink observatory, Dublin Institute for Advanced
Studies, Dublin 15, Ireland}\ec

\vs\vs

\begin{quotation}

\no {\bf Abstract}\hs An attempt to answer the question ``Can
we observe galaxies that recede faster than light'' led to
a re-examination of the notions of time, distance, velocity
and redshift as they occur in newtonian physics, special
relativity, general relativity and cosmology. A number of
misconceptions were uncovered. It was found  that, once freed
of preconceptions of special relativity the above question
is easily and unequivocally answered. 

\end{quotation}

\vspace{1cm}

\no Over the past fifteen years or so I have been a student
of cosmology. I was puzzled by the following question:
according to the celebrated Hubble's Law that says velocity
of recession is proportional to distance, there must come a
distance so large that a galaxy located there is receding at
the speed of light.  What then happens to the photon the
galaxy emits in our direction ?  Will that photon ever reach
us ?  Or if it does reach us, will its redshift be infinite
and hence it will make zero impact ?  In that case, that
distance must be marking the boundary of our {\em
observable\/} universe; is that distance what is known as
horizon ?  Or has the Hubble Law broken down long before that
point ? Or have we missed the point entirely and forgotten
that velocity at a distant point is simply not defined in
General Relativity ?  I eventually worked out the answers and
these were given in my (1997) paper. And it dawned on me that
all the above ``questions'' are wrong questions, prompted by
what I call the ``Shadows of Special Relativity''. Once it is
recognized that modern cosmology, dubbed ``relativistic'', is
based {\em not\/} on Special Relativity, but on General
Relativity, and {\em not\/} on General Relativity alone, {\em
but also on\/} the Cosmological Principle, the answers emerge
with no more than first-year college mathematics.  The
present essay is a systematic clearing-up of all the dead wood
and misconceptions encountered on the way.  I have not
refrained from using non-technical language, homely
analogies, formal comparisons whenever these contribute to
undesratanding.

\vs\vs   
\bc {\large\bf 1. Event-Intervals and Observers} \ec

Physics deals with events. But not {\em individaul\/} events;
rather, what matters is the {\it interval\/} between
specified events. An {\em observer\/} O specifies a given
event by three (newtonian) spatial coordinates $(x, y, z)$,
and one (newtonian) time coordinate $t$. Consider a second
observer O$'$, fixed at ($X_0, Y_0, Z_0 $) with respect to O,
and whose spatial axes O$'x'y'z'$ are parallel to Oxyz. Let O$'$
specify the same event by $(x', y', z', t')$. It is usually
presumed that according to Newton we should have simply
$t'=t$. But that is {\em not\/} the essence of newtonian
time. Newton characterises time as that which by its own
nature {\em flows\/} evenly without reference to anything
else, which means that so long as the clock {\em rates\/} of
O and O$'$ are the same, their {\em readings\/} can differ by a
constant amount, $T_0$, say. For this event, then, we have 
%% (1)
\bea 
\left. \begin{array}{l}
    x=x'+X_0,\hs y=y'+Y_0,\hs z=z'+Z_0\,, \\ 
    t=t'+T_0\,.
\end{array} \right\}
\eea  
Now consider two events E\dn{1} and E\dn{2}, with their
respective specifications by O and O$'$:      
%% (2)
\bea  
\left. \begin{array}{l}
      {\rm E\dn{1}} \equiv (x_1, y_1, z_1, t_1) \equiv 
	(x'_1, y'_1, z'_1, t'_1)\;, \\ 
      {\rm E\dn{2}} \equiv (x_2, y_2, z_2, t_2) \equiv 
	(x'_2, y'_2, z'_2, t'_2)\;.
\end{array} \right\}
\eea
\no For each of the two events, relation (1) holds, hence, for
the interval between the two events we have, writing $\Delta
x$ for $x_2-x_1$ etc., 
%%(3)
\bea
\left. \begin{array}{l}
   \Delta x=\Delta x',\hs \Delta y=\Delta y',\hs \Delta z=\Delta
	z'\,, \\ 
   \Delta t=\Delta t'\;.
\end{array} \right\}
\eea 
\no These equations are what Newton would have for two
observers at rest each to the other.

What happens if O$'$ has a velocity $v$ relative to O along the
x-axis ?  Note, so far, only newtonian concepts of time,
distance and velocity are involved. Continuing within the
newtonian framework, we say that in that case, we have
%% (4)
\bea
\left. \begin{array}{lcr}
   \Delta x   & = &   \Delta x'  +  v \Delta t'\;, \\ 
   \Delta t   & = &                   \Delta t'\;. 
\end{array} \right\}
\eea
\no and $\Delta y=\Delta y', \Delta z=\Delta z'$. From this 
point on, these two trivial relations will generally be left
implicit. Note in the first of the two equations, Newton
would have written the additional term due to $v$ as $v\Delta
t$ and left the second equation implicit:  I have written out the
second and used it to put the first in the form shown, in order
to contrast it with what is forthcoming.

Eqs.\ (4), then, are the coordinate (difference)
transformation according to Newton for two observers in
relative (uniform) motion.

\vs\vs
\bc{\large\bf 2. Special Relativity: Some Common Misconceptions}\ec

Now consider a photon moving in the $x$-, or equivalently, the
$x'$-direction and let us identify the two events E\dn{1} and
E\dn{2} with the photon occupying two specific points at two 
instants of time. Then, if we write $c$ (=$\Delta x/\Delta
t$) for its speed measured by O, and $c'$ ($=\Delta x'
/\Delta t'$) for its speed measured by O$'$, then by dividing
the first equation in (4) by the second, we obtain
%% (5)
\be      c = c' + v .
\ee
This is contrary to Galileo's principle that no experiments
can detect a uniform motion. And Einstein was such a firm
believer in the Galilean principle that he would rather have
the transformation (4) changed so as to have $c=c'$ for all
observers in relative (uniform) motion. The result is the 
famous Lorentz transformation, which is most neatly written
when expressed in some $c=1$ units (e.g., all time-intervals in 
years and all distances in light-years), thus:
%% (6)
\bea
\left. \begin{array}{l}
     \Delta x = \gam\Delta x' + \bet\gam\Delta t'\;, \\ 
     \Delta t = \bet\gam\Delta x' + \gam\Delta t'\;.
\end{array} \right\} \eea 
\no where $\bet=v/c, \gam=1/\sqrt{1-\bet^2}$.

\vn\underline{From this point on, $c=1$ units will always be
assumed}\hs
Readers may have noticed that most textbooks write the
Lorentz transformation differently, that they would 
drop the $\Delta$'s and write $x$ for $\Delta x$, etc., thus: 
%% (7)
\bea
\left. \begin{array}{l}
       x = \gam  x' + \bet\gam  t'\;, \\ 
       t = \bet\gam  x' + \gam  t'\;.
\end{array} \right\} 
\eea 
This form is what the form (6) reduces to in the special case
where O and O$'$ both assign the coordinate values $(0, 0, 0,
0)$ to the event E\dn{1}. It is my belief that it is the
failure to recognize that the form (7) represents a special
case and is not {\em always\/} applicable that is the cause
for all the confusion surrounding the so-called ``twin
paradox'' (Kiang 1992).  
%%%%%%%%%%% 
\footnote {The differential ageing between two twins, one a
stay-at-home, one a space traveller, can be illustrated with
the following simple numbers.  On their common 20th birthday,
twin A stays behind while twin B goes out into space at speed
$0.6c$.  On B's 30th birthday he turns round and heads back
at the same speed $0.6c$. He reaches home on his 40th
birthday, just to find A celebrating {\em his\/} 45th
birthday. Between parting and re-unison (two specific
events), B has aged 20 years and A, 25 years.  This result
follows unequivocally from the form (6), when we recognize
that, here, {\em three\/} inertial frames (or newtonian
observers in relative (uniform) motion) are involved, No.\,1
for A, No.\,2 for B going out, No.\,3 for B coming back. If
we use the form (7) {\em both\/} for the transformation
between No.\,1 and No.\,2 {\em and\/} for that between No.\,1
and No.\,3, then we could multiply paradoxes {\em ad
infinitum\/}. For details, see Kiang (1992)}.  
%%%%%%%%%%%%

It is interesting to compare the Lorentz transofrmation (6)
with the Newton transformation (4). The spatial interval 
$\Delta x$ is {\em qualitatively\/} the same in both (being
dependent on both $\Delta x'$ and $\Delta t'$), but is {\em
quantitatively\/} different (the coeeficients are different),
while the time interval $\Delta t$ is even {\em
qualitatively\/} different: in Newton, $\Delta t$ simply
equals $\Delta t'$; in Lorentz, $\Delta t$ depends on both
$\Delta t'$ and $\Delta x'$ (and in the same way as $\Delta
x$ depends on $\Delta x'$ and $\Delta t'$). 

Eqs.\,(6) imply $(\Delta t)^2 - (\Delta x)^2 = (\Delta t')^2
- (\Delta x')^2 $. Then, incorporating the two triavial
relations spelled out above, we can subsume the whole family
of Eqs.\,(6) for all values of $v$ between 0 and 1 by one 
single statement, namely, the finite ``spacetime interval''
$\Delta s$ defined by 
%%(8)
\be 
    (\Delta s)^2 = (\Delta t)^2 -
		[(\Delta x)^2 + (\Delta y)^2 + (\Delta z)^2] 
\ee
\no is invariant (has the same value) for all observers in
relative (uniform) motion. 

\vn{\bf 2.1~~Minkowski} noted that if we put $\tau =it$ in
the above formula, then $\tau$ would be formally
indinstinguishable from $x, y$ or $z$, so he wrote his famous
words,

\begin{quotation}
``Henceforth space by itself and time by itself, are doomed
to fade away into mere shadows, and only a union of the two
will preserve an independent reality'' (quoted in Taylor and
Wheeler 1963 p.37). 
\end{quotation}

These words are somethimes taken (or mistaken) to mean that 
space and time must now be considered as indistinguishable.
That time and space are indistinguishable is generally true
in the next stage of the theoretical development---Einstein's
General Theory of Relativity (GR), but here at the stage of
his Special Theory of Relativity (SR), it is not true. In SR,
for a given observer, space intervals are still space
intervals, time intervals are still time intervals: the
observer conceives space and time exactly as Newton did.
Only in SR, for two observers in relative (uniform) motion
measuring the separation between the same pair of events, the
time or spatial {\em measure\/} of one is {\em each\/} related to
{\em both\/} the time and spatial {\em measures\/} of the other.

\vs\vs
\bc{\large\bf 3. General Relativity: Merging of Time and Space}
\ec

Gravity is not considered in SR. It was to incorporate
gravity that Einstein developed his GR, with the guiding
principle that GR reduces to SR at the local limit. In GR, an
event is specified by four {\em generalised\/} coordinates,
$(\xi^\mu, \mu=0,1,2,3)$ and, instead of {\em finite\/}
intervals $\Delta \xi^{\mu}$, GR deals with {\em
infinitesimal\/} intervals $d\xi^{\mu}$. Corresponding to
(8), GR has as invraint the quantity, called ``the metric''
or the ``line-element'' $ds$, defined by       
%%(9)
\be
	(ds)^2 = g_{\mu\nu}\,d\xi^\mu\,d\xi^\nu\;, 
\ee 
\no where a repeated index implies summation over (0,1,2,3),
and $g_{\mu\nu}$ ($g_{\mu\nu}=g_{\nu\mu}$ for $\mu\neq\nu$),
are ten independent functions of $\xi^\mu$, that inform the
local gravitational field. 

It is instructive to compre the two invariant forms, GR's (9)
and SR's (8), or rather, the latter's implied infinitesimal
form,      
%(10)
\be
	(ds)^2 = (dt)^2 - [(dx)^2 + (dy)^2 + (dz)^2] 
\ee
\no We see how much simpler (10) is: it contains no mixed
terms such as ``$dt\,dx$'' and the four non-zero $g_{\mu\nu}$
(partiuclarly that factors $(dt)^2$) are all constants. 

If in (10) we replace the rectangular coordintes $(x, y, z)$
by the usual spherical coordinates $(r, \theta, \phi)$ we get
%%(11)
\be 
    (ds)^2 = (dt)^2 - 
      \{(d r)^2+r^2[(d\theta)^2+\sin^2\theta(d\phi)^2]\}\,. 
\ee 
\no This form of the Minkowski metric will be used below for
comparison purposes.

\vn{\bf 3.1 Condition for a Universal Time in GR} 

At a given event, GR must reduce to SR, so at a given event,
labelled 1, it is always possible for us to distinguish, say,
$\xi^0$ as time, $t$.  Can the same identification be made at
another event 2 ?  Now, in GR, this means we have to see what
happens when we integrate $ds$ between 1 and 2 along the path
$\xi^1=\xi^2=\xi^3=$ const., or 
$ds=\sqrt{g_{00}}\,d\xi^0$. Hence

(i) if $g_{00}={\rm const.}$ at all events, then $\xi^0$ is
indeed universal time; or,

(ii) if at all events, $g_{00}$ is a function only of
$\xi^0$, $f(\xi^0)$ say, then we can take $\int
f(\xi^0)d\xi^0$ as universal time.

These conditions are so restrictive that, in the general
case, we do not expect to be able to say which one of the
four coordinates is time: time and space {\it are\/} mixed up
in a single, four-dimensional continuum, spacetime. 

Yet in one of the most important applications of GR, namely,
cosmology, time and space separate out;---and in a most
characteristic way too. The time that emerges is universal
time and the space, is of a kind hitherto undreamed of. 

\vs\vs
\bc{\large\bf 4. Relativistic Cosmology: Space Evolves in 
Time}\ec   

That this remarkable situation emerges in relativistic cosmology
is because the latter is not just General Relativity: it is 
General Relativity plus the Cosmological Principle (CP),
which says the unvierse is homogeneous and isotropic and
looks the same to all observers.
%%%%%%%%%%%%
\footnote{The last part is usually separately known as the
Copernican Principle}.
%%%%%%%%%%%%
The CP implies that the three-dimensional physical space must
either be static, or expanding uniformly or contracting
uniformly, in complete analogy with the corresponding
behaviour of the two-dimensional surface of a balloon that is
readily visualizable.  Observations (mainly associated with
the name of Edwin Hubble) then tell us that the universe is
expanding, rather than static or contracting.    

The solution of Einstein's field equation (the ``master''
equation of GR) under the CP is the
Friedmann-Robertson-Walker (FRW) metric,
%%(12)
\be 
    (ds)^2 = (dt)^2 - 
      R^2(t)\{(dr)^2+r^2[(d\theta)^2+\sin^2\theta(d\phi)^2]\}\,.
\ee 
\no It should be said that this is only one (the simplest) of
three possible forms of the FRW metric: it corresponds to the
case of {\em flat\/} space ($k=0$). If space is closed
($k=+1$) or open ($k=-1$), then the $r^2$ factoring the
square brackets should be replaced by 
      \[ k a_0^2 \sin^2(r/\sqrt{k}a_0)\;, \hs (k=\pm 1),  \] 
\no containing an additional free parameter $a_0$, the radius
of curvature of space at the present epoch (Longair 1984,
p.292). This way of specifying the FRW metric has the
merit that, irrespective of space curvature, $r$ is always
the (comoving) radial coordinate itself;---and we shall be
concerned exclusively with radial motions 
(e.g., Eq\,(17)).
%%%%%%%%%%     
\footnote {Many authors (e.g., Felten et al.\ 1986) prefer to
use the Schwarschild radial coordinate, $r_1$ say, such that
$2\pi r_1$ is the comoving circumference (Misner et al.\
1970, p.723), and replace the $(dr)^2$ in (12) by
$(dr_1)^2/[1-k(r_1/a_0)^2]$.} 
%%%%%%%%%%%

Let us now compare the FRW metric (12) with the ``Minkowski''
metric of SR, (11). Formally they have one feature in common;
and there is also one all-important difference. 

In both, we have a universal time $t$, as evidenced
by the term $(dt)^2$ and by the absence of mixed terms such
as $dt\,d\phi$. (cf.\ 3.1).

The difference is in what happens to the radial line element
$dr$ in the two cases. In the Minkowski, $dr$ stands by
itself; in the FRW, it is factored by $R(t)$, a function of
the time $t$. 

Recall that in SR, both time and space are newtonian (cf.
2.1). Hence we have that in relativistic cosmology, time is
still newtonian, 
%%%%%%%%%%%%%%%%%
\footnote{The Cosmological Principle that the universe must
look the same to all observers is meaningless if the universe
is evolving, unless we qualify the ``look the same'' with ``at
a given time''. Then a given appearance or state of the
universe as measured by, say, the rate of expansion and/or
the mean density, must be associated with a particular
instant of time, a particular clock {\em reading\/}. So in
regard to time, cosmology has out-newtoned Newton: not only
do all clocks run at the same rate, they actually give the
same reading to a given event.}
%%%%%%%%%%%%%%%%%
but space is no longer so: all spatial intervals now depend
on a single, universal function of time, $R(t)$,  called
``scale factor''. The scale factor describes how all
distances evolve in time, it is the essence of an evolving
universe.

\vn{\bf 4.1~~Distance and Coordinate-Distance}

Central to a universe in which space evolves uniformly in
time are two distnct notions, ``distance'' and
``coordinate-distance''
%%%%%%%%%%%%%%
\footnote{``Distance'' is what is known as ``proper
distance'' in GR, and ``coordinate-distance'' is sometimes
called ``comoving distance''}.
%%%%%%%%%%%%%%
So central are they that they deserve to be introduced by 
a homely analogy.

Let us imagine that our earth is expanding, that its radius
$a$ is a function of the time $t$: $a=a(t)$. Let us denote
the present epoch by $t_0$. ($a(t_0)=6371$\,km).  Let us
suppose that we the observer are located at the North Pole.
Now let us consider the distance of a place such as Datong
from us. Let the co-latitude of Datong be denoted by $\chi$
(=50\dg). Let us denote its present distance by $r$: we have 
    \[ r = a(t_0) \cdot \chi\hs, \]
and let us denote its distance at time $t$ by $D$: we have 
    \[ D = a(t) \cdot \chi \hs. \]
Now we introduce the scale factor $R(t)$ defined as 
    \[ R(t) = a(t)/a(t_0)\;, \hs (R(t_0)\equiv 1)\hs. \]
Then the previous expression can be re-written as 
%% (13)
\be
     D = R(t) \cdot r \hs, 
\ee
\no and this is the basic relation between the ``distance''
$D$ on one hand and the ``coordinate distance'' $r$ 
and the (normalised) scale factor $R(t)$ on the other.  The
distance from the observer to any point fixed to the earth,
$D$, depends on time thorugh $R(t)$, and the value of $D$ at
some standard time at which $R=1$, is the coordinate-distance
of that point, $r$: each point fixed to the earth is
associated with a fixed value of $r$. For Datong,
$r=6371\x(50\dg$ in radians)$=5560$\,km; for Guadacanal ($\chi
=100\dg$), $r=11120$\,km. Guadalcanal is twice as far as
Datong {\em now\/}, and will always be so, however
the earth expands (or shrinks). 

The situation in the expanding universe is exactly the same.
All we have to do is to identify the $r$ and $R(t)$ in (13)
with the $r$ and $R(t)$ in the FRW metric (12), and use (13)
to define a distance $D$ in the universe. The distance, $D$,
from us to a particular point in the evolving space, then, is
a function of {\em two\/} varaibles, the universal time $t$
through the universal function $R(t)$, and the
coordinate-distance $r$ of the particular point under
discussion:  $D=D(t,r)$. The particular point {\it may or may
not} happen to be occupied by a galaxy, a quasar, a speck of
dust, a source of radiation, or a passing photon: its
distance is always equal to the product of its
coordinate-distance and the universal scale factor.

At any given time $t$, $D$ is simply proportional to $r$: just
like Datong and Guadalcanal, if the distance of quasar A at
the present epoch is 1 Gly (one billion light-years), and
that of quasar B is 2 Gly, then B will always be, and has 
always been, twice as far as A, whatever the form $R(t)$
in the future or in the past. 

For a particular point in space with coordinate-distance
$r$, its velocity of recession
%%%%%%%
\footnote{From now on we shall always talk of ``recession'',
with the understanding that it could be negative},
%%%%%%%
$v^{\ast}$, is given by the partial derivative of $D$ with
respect to $t$, at fixed $r$: 
%%(14) 
\be  
	v^{\ast} = \dot{R}(t)\cdot r \;.  
\ee 
\no This is, in fact, the primitive and and a more
practically useful form of Hubble's Law. The usual form of
Hubble's Law is 
%% (15)
\be  
    v^{\ast}= H(t)\cdot D\;, \hs (H(t)\equiv\dot{R}(t)/R(t))
\ee

\no obtained by eliminating $r$ between (14) and (13), and
introducing the Hubble parameter $H(t)$ as shown.

\vn{\bf 4.2 Status of Hubble's Law}

A common misunderstanding surrounds Hubble's Law.  It is
often thought that Hubble's Law comes directly from Hubble's
observations of the galaxies. Not so. Hubble, of course, did
not observe {\it directly\/} the velocities and distances of
the galaxies.  What he observed was their redshifts and
apparent magnitudes and he found a correlation between the
two.  Then, interpreting the redshifts as Doppler shifts, and
the apparent magnitudes as a distance effect, the observed
linear regression of redshift on apparent magnitude is
converted into a linear relation between velocity of
recession and distance, and the latter has come to be
popularly known as Hubble's Law. This law is thus based on an
interpretation of an empricial correlation, and as such,
cannot be an exact law. (In fact, Hubble's original data had
a very large scatter (Hubble 1936), but whatever refinement
later observations have brought about, an emiprical relation
can never imply an exact relation). Considered this way, it
would be legitimate to say that, possibly, or even probably,
the linearity between velocity and distance in the law is
only approximate, valid only  for small distances, and that
as the distance increases indefinitely, the velocity of
recession may approach the speed of light only
asymptotically. That would neatly reconcile the extrapolation
of this empirically-based law with the requirement of SR that
nothing can move faster than light.  (And after Hiroshima who
can doubt the truth of SR ?)

However, this is not what is understood as Hubble's Law in
thoeretical cosmology. Modern cosmological models come
directly from GR and the CP, and from the latter comes the
basic relation (13), and hence, the distance-velocity
relation (14) or (15).  It is this {\em strictly linear\/},
observation-independent, relation between distance and
velocity of recession (with, however, a time-dependent
``constant'' of proportionality, the Huble parameter
$H=H(t)$) that is now understood as Hubble's Law.  From this
perspective the significance of Hubble's and his successors'
observations of the galaxies is just to tell us that the sign
of the present value of $H$ or $\dot{R}$ is positive, rather
than zero or negative, and to tell us what that positive
value is.  
%%%%%%%% 
\footnote{For orientation purposes, a nice round number of
the Hubble time, $H_0^{-1}$, is 20 Gyr (20 billion years). 
Latest estimates would put it some 30\% lower.}. 
%%%%%%%% 
According to the Hubble's Law (14), then, the velocity of
recession at any time $t$ is strictly proportional to the
coordinate-distance $r$.  If the universe is spatially open
($k=-1$) or flat ($k=0$), then $r$ can be indefinitely large,
and so can the velocity of recession. If space is closed
($k=+1$), then $r$ and hence velocity of recession, has a
finite upper bound, but for any closed model at all
consistent with the observtions, recession velocities greater
than $c$ will have existed since the ``beginning'' up to a
certain far point in future. For details and numerical
illustrations see Kiang (1997).

\underline{A {\em Caveat\/} on the term `superluminal'}$\;$
Recession velocities greater than the speed of light are
sometimes referred to as ``superluminal'' velocities. They
should be carefully distinguished from the ``superluminal
velocities'' (often associated with ``tachyons'') discussed
by experimental and theoretical physicists. To add to the
confusion, ``superluminal jets'' are said to have been
observed in many radio sources. In this last case, the
``superluminality'' is a purely apparent effect  
%%%%%%% 
\footnote{If a jet leaves a source at a speed $v$ and at an
angle $\psi$ to the direction towards the observer, then its
apparent velocity, $v_{\rm app}$, defined as the ratio of its
transverse linear separation (got from its observed angular
separation and any assumed large distance) to the observed
time-interval, is, $v_{\rm app}=v\sin\psi/(1-v\cos\psi)$. The
derivation is entirely newtonian, assuming only a constant
light speed.  Then for a $v$ under 1, we can get $v_{\rm
app}$ greater than 1. It is often thought that in addition,
$\psi$ has to be small; this is not true, in fact, for a
given $v$, $v_{\rm app}$ maixmizes at $\psi=45\dg$. In the
astronomical context, however, small $\psi$ is necessary for
the jet to be sufficiently boosted relativistically to become
visible (Rees 1967).}.       
%%%%%%%%

The point in space where the recession velocity equals $c$ is
nothing special. That we should have thought it special is
because we have been psychologically conditioned to think
always in terms of SR. It is high time to recognize the {\it
unhelpful\/} character of SR thinking when working on
cosmology.

In SR, $v=c$ is associated with infinite redshift, synonymous
with observable limit. Now, recognising that SR is not valid
in cosmology, the question  ``Can we observe a galaxy that
recedes at the light speed ?'' can no longer be dismissed with
a categorical ``Of Course Not'': the question has become
non-trivial and interesting. And entirely tractable, as it
turned out. 

Nor can the question be dismissed as meaningless, on a higher
level, by saying velocity at a distant point is not defined
in GR. Cosmolgy is not just GR, it is GR plus the CP, and
thanks to the latter, finite distances, and velocities at
distant points, all referring to particular instants of
time, are all perfectly defined.

\vn{\bf 4.3  How Photons Move in Expanding Space }

Let us go back to the analogy of the expanding earth. Let us
now suppose there is a special race of ants constantly
crawling on the surface of this earth. The ants are special
in that they always crawl at the same {\em ground\/} speed
{$\kappa$}, regardless of what the ground is doing. Now
consider a certain instant $t$ and a certain location
specified by co-latitude $\chi$, such that, at that instant
$t$, that place is receceding from the North Pole at speed
$\kappa$:  $\dot{a}(t)\chi=\kappa$. We now ask, ``Suppose an
ant starts there and then to crwal towards us at the North
Pole, will it ever reach us ?'' It is obvious that the answer
must simply depend on the given form of $a(t)$, and can be
found using elementary calculus.

But this is exactly the same as the ``cosmological'' question
we are asking, once we identify the ants with photons and
$\kappa$ with $c$. So the answer to our question simply
depends on the form of $R(t)$, and can be got just as easily.
 
Now we leave the analogy and work out the cosmological
question in full accordance with GR. The equation of motion
for a photon in GR is given by $ds=0$. Now, we are only
interested in photons moving in the radial directin, so the
FWR metric (12) is simplified to read
%%%%%%(16)
\be  
	(ds)^2 = (dt)^2 - R^2(t)\cdot (dr)^2  
\ee
\no and $ds=0$ is simply,
%%%%%%%(17)
\be
	dr = \mp dt/R(t) 
\ee
\no  with the minus sign for incoming photon and the plus
sign for outgoing photon. 

This equation is the basis for answering the question that
started the present enquiry. It also provides definitions of
horizon and redshift in cosmology. It turned out that the
distance at which $v^{\ast}=1$ has nothing to do with
horizon, nor anything to do with infinite redshift.

\vn\underline{4.3.1~~Can we observe galaxies that recede fast
than light ?}

Starting with Eq.\,(17) for the incoming photon and the
primitive form of Hubble's Law (14), and using no more than
elementary calculus, I was able to answer this question
completely  (Kiang 1997). The answer is as follows. For the
steady-state model ($R(t)= \exp (t), -\infty<t<+\infty$, in
Hubble units
%%%%%%%%%%%
\footnote{We use Hubble units when we express all time
intervals in units of the Hubble time, $H_0^{-1}$ and all
distances in units of the Hubble radius, $cH_0^{-1}$. Hubble
units are one example of $c=1$ units, and the latter are a
sub-set of ``normalised units''.  For a general account of
``normalised units'', see Kiang (1987)}
%%%%%%%%%%%
), the answer is ``No''. For all the three varieties ($k=0,
\pm 1$) of the big bang model, the answer is ``Yes'';
moreover, we can say, for example, that for the $k=0$
``standard'' model, all quasars we now observe having
redshifts greater than 1.25 (we now know thousands of these)
have the property that, at the time of emission of the
photon that now reaches us, they were receding faster than
$c$. More recently (Kiang 2003), I have generalised the
answer to, ``Yes, if the universal expansion started with a
singularity; No, if it started infinitely slowly from a finite
size including zero;---this happens when the model contains a
cosmological constant at a certain critical value (Bondi 1960,  
p.\ 82, Felten and Isaacman 1986)''.

\vn{\underline{4.3.2~~Horizons}

Starting with Eq (17) with the plus sign, and integrating
from $t_{\rm begin}$ (=0, or $-\infty$ as the case may be) to
any given epoch $t$ results in the (particle) horizon (size
of the observable universe) at the time $t$ (Kiang 1991). If
the universe had a finite past, then the horizon is finite.
And obviously, this finite horizon has nothing to do with the
distance at which $v^{\ast}$ assumes any particular value
including 1. 

Integrating the same equation from a given epoch $t$ to
$t_{\rm end}$ ($=+\infty$, or some finite value, as the case
may be), results in the ``event horizon'' at $t$, which 
does not directly concern the present enquiry (Kiang 1997)
%%%%%%%%
\footnote{Confusingly, researchers on formation of large
scale structures now use yet a third ``horizon'', which turns
out to be just the Hubble radius at the current epoch,
$c/H(t)$}.
%%%%%%%% 

The important thing to note here is that horizon, or the
limit of observability in the universe, is a certain simple
function of $R(t)$. It has nothing to do with infinite
redshift, as we shall see presently. That the two should be
linked is another misleading hang-over from SR.

\vn{\bf 4.4 Cosmological Redshift. Doppler Redshift }  

The equation of the incoming photon, Eq.\,(17) with the minus
sign, is again used to derive the formula for the redshift in
cosmology. The formula is (see e.g., Bondi 1960, p.106), 
%%%%%%%%(18) 
\be    1+z = \frac{1}{R(t_{\rm em})}\;, \hs 
			(R(t_{\rm obs})\equiv 1)\;,
\ee
\no where $t_{\rm em}$ is the time of emission of the photon,
and $t_{\rm obs}$ is the time of its observation, identified
with the present epoch $t_0$. Thus, cosmological redshift has
nothing to do any velocity; it simply depends on the value of
the scale factor at the epoch of emission; it is completely
different from Doppler redshift (whether classical or
relativistic), which is above all else a function of the
radial velocity of the source with respect to the observer.

Of course, for a given model, i.e., a known $R(t)$, it is easy
to work out a relation between the redshift $z$ and the
velocity of recession of the source {\em at a specified
moment\/}, e.g., the time of emission $t_{\rm em}$, or the
time of observation (the present epoch), $t_0$. For the
simplest big bang model (the standard model), with
$R(t)=(t/t_0)^{2/3}$, I derived (Kiang 1995), 
%%%%%%% (19)
\bea 
\left. \begin{array}{l}
            v^{\ast} (t_{\rm em}) = 2 (\sqrt{1+z}-1)\,,\\
	    v^{\ast} (t_0) = 2 (1-1/\sqrt{1+z})\,.
\end{array} \right\}
\eea  
However, it could never be over-emphasised that these
particular $v^{\ast}-z$ relations are altogether different in
character from the relativistic Doppler redshift formula
%%%%%%%%
\footnote{The formula (20) is for purely radial motion. For 
motion at angle $\theta$ to the positive radial direction,
the formula is $1+z= (1+v\cos\theta)/\sqrt{1-v^2}$.},  
%%%%%%%% (20)
\be
        1 + z = \sqrt{1+v} / \sqrt{1-v}  \;.
\ee

\no The formal difference is that each $v^{\ast}-z$ relation
is specific to a cosmological model and for a particular
characteristic time, whereas the Doppler formula is quite
general. But the fundamental difference lies between $v$ and
$v^{\ast}$ in their meaning:  the $v$ in the Doppler formula
refers to motion of objects {\em in\/} space which itself is
tacitly assumed to be {\em fixed\/}, whereas the $v^{\ast}$
in (19) refers to motion of objects that are merely being
carried along by a space, which iteself is {\em expanding\/}.
Equation (13), expressing the 
expansion, is basic to cosmology; it is alien to Special
Relativity. We would only be misleading ourselves if we apply
consequences and deductions of SR, deductions such as
``nothing can move faster than light'', or ``$`v=c$ implies
infinite redshift'', to an expanding universe.

\vn\underline{4.4.1 The Redshift of Superluminal Jet}. 

Superluminal jets (cf.\ the {\em Caveat\/} in 4.2) offer an
instance where we are forced to consider simultaneously the
two types of redshift, cosmological and Doppler. The problem
can be formulated as follows. A radio source with observed
redshift $z_{\rm source}$ has an (apparently) superluminal
jet, which is supposed to be due to the jet moving at
velocity $v_{\rm jet}\siml 1$ relative to the source at a
small angle $\psi$ towards us. What would the redshift of the
jet ($z_{\rm jet}$) be, if observable ? The answer I found
(Kiang 2003) is that $z_{\rm jet}$ is to be given by 
$1+z_{\rm jet} = (1+z_{\rm source}) (1+z_{\rm Doppler})$,
where the last factor is to be calculated from the
relativistic Doppler formula for the given $v$ and
$\theta=180\dg-\psi$, shown in the footnote to
Eq.\,(20).

\vn{\bf 4.5~~Recession Velocity. Peculiar Velocity }

The versatile Eq.\,(17) describes the time variation of the
corrdinate-distance $r$ of the photon moving in the radial
direction.  What about the time rate of its (proper) distance
$D$\,?  From the definition of $D$ at (13), we have,
immediately,
%%%%(21)
\be 
    \frac{dD}{dt} = r\cdot\frac{dR}{dt} + R\cdot\frac{dr}{dt}
\ee
\no The first term on the right is,  accrding to (14), just
the recession velocity at the current point, $v^{\ast}$, and
the second reduces to $\mp 1$ on using (17). Hence the
``total'' velocity of the photon (relative to us), 
at time $t$, at distance $D$, is
%%%%%(22)
\be  
       v_{\rm photon} = v^{\ast} \mp 1 \;.
\ee
This equation says that the local velocity of the photon
($\mp 1$) is simply compounded with the recession velocity 
$v^{\ast}$ in the old newtonian, pre-relativity manner (Kiang
1997). And there is nothing remarkable about that, once we
are rid of Special Relativity preconceptions. 

Formula (22) is a particular case of the more general formula
for the ``total'' velocity (Davis and Lineweaver, 2000), 
%%%%%%%%%
%%%% (23)
\be
	v_{\rm total} = v^{\ast} + v_{\rm peculiar} \;,
\ee     
particularised to the case $v_{\rm peculiar}=\mp 1$. 

Another particularization would be the case of a 
relativistic jet coming out of a distant radio source,
where the recession velcity would be referring to the source
and the peculiar velocity is the radial component of 
the relative velocity of the jet with respect to the source.   

A third particularization would be the case of a ``comoving
source'' where $v_{\rm peculiar}$ is identically zero, and
the total velocity is just the recession velocity.

Recession velocity and peculiar velocity are different
animals. The recession velocity should not be regarded as the
property of a source; rather, it should be considered as the 
property of the point of space in question, whether that
point happens to be occupied by a source, a passing photon,
or nothing at all.  The peculiar velocity, on the other hand,
must have reference to a material object including photon. 
The recession velocity is not the kind of velocity considered
in SR, hence it is not subject to any of the laws of SR; the
peculiar velocity {\em is\/}, and so {\em is\/} subject to
all its laws: it must never exceed the speed of light and it
gives rise to Doppler shift (Cf. 4.4.1). The recession
velcoity is always in the radial direction, whereas the
peculiar velocity can be in any direction;---what appears in
Eq.\,(23) is its radial component. After detailing all these 
essential differences, we almost marvel at the simply way
they combine to give a total velocity. The simplicity comes
from the underlying simplicity of the expanding space, as
expressed by the Hubble Law. The two velocities are additive
simply because they are just the two terms in the
total differential of a function of two variables.

\vs\vs
\bc {\large\bf 5~~Shadows of Special Relativity} \ec

After Hiroshima, who of us can doubt the truth of Special
Relativity ?  So we take it for granted that nothing can move
faster than light, that $v=c$ means infinite mass, infinite
redshift, some absolute physical barrier, etc.. Then Hubble's
Law comes along and seems to say that there {\em are\/}
galaxies that recede faster than light.  How can we square
that with SR ? So we consider a series of ways out.  These
are the ``leading questions'' at various levels spelt out in
the introductory paragraph of this essay.

One by one these escape routes are proved untenable. They are
untenable because we shouldn't have considered them in the
first place. The recession velocities of galaxies are, as it
were, outside the remit of SR, the expanding space in
cosmolgy is not the fixed space of SR, so why {\em
shouldn't\/} there be contradictions between SR and cosmology
?

Once the shadows of SR are cast away, the following broad 
features of our universe emerge easily enough. The difficulty
has not been the mathematics, but the psychology.

1. At any given time the recession velocity is strcitly
proportinal to the distance and there are certainly galaxies
that recede faster than light.

2. Galaxies receding faster than light are observable if the
universal expansion started with a bang at a finite past
epoch; they are not observable, if the expansion started
infinitely slowly.

3. Galaxies that recede at the speed of light are NOT
associated with infinite redshift. On the contrary, infinte
redshift in cosmology is associatged with the {\it time\/} of 
the bang, if it exists, when the size of the universe is zero.

4. Horizon in the sense of a boundary to the observable
region exists if the universe started a finite time ago
%%%%%%%%%
\footnote{This statement also applies to a static universe.
Horizon is certainly {\it not\/} a privileged property of
expanding universes; any universe, static or expanding, has
a horizon, so long as it has a finite past. In fact, for a
static universe, the horizon is simply $ct$}
%%%%%%%%%
and then its value depends only on the form of the scale
factor and has nothing to do with redshift.

It has been a personal journey of discovery. On the way I
have identified a number of common misconceptions about time
and space.  I have pinpointed the origin of the twin paradox
in the usual form of the Lorentz transformation given in
textbooks. I have recognized the role played by the
Cosmological Principle in lifting cosmology out of the
general indeterminateness of General Relativity by providing
it first with a universal time, then with well-defined
notions of finite distances at a given time, and of
indefinitely large recession velocities (regardless of the 
speed of light) at indefinitely large distances.


\begin{thebibliography}{99}

\bibitem{} Bondi. H, 1966, Cosmology, Cambridge University
Press, p.\,82, p.\,106.

\bibitem{} Davis, T. M. and Lineweaver, C. H., 2000,
presented at Cosmology and Particle Physics 2000, Verbier,
Switzerland, appearing in American Institute of Physics
conference proceedings Volume 555, editors Durrer, R.,
Garcia-Bellido, J. \& Shaposhnikov, M., AIP, New York.

\bibitem{} Felten J. E., and Isaacman, 1986, Rev. Mod.
Phys., 58/3: 689-698.
				       
\bibitem{} Hubble E., 1936, The Realm of the Nebulae, Oxford
University Press.

\bibitem{} Kiang T., 1987, Quaterly Journal of the Royal
Astronomical Society, 28: 456-471. Reprinted as Contributions
from the Dunsink Observatory No.\,27.    

\bibitem{} Kiang T., 1991, Acta Astrophysica Sinica (Tiantiwuli 
Xuebao), 11/3, 197-212. Reprinted as Contributions from the
Dunsink Observatory No.\,31.     

\bibitem{} Kiang T., 1992, Irish Astron, J., 20: 201-206.    

\bibitem{} Kiang T., 1995, Irish Astron, J., 22/2: 159-163.

\bibitem{} Kiang T., 1997, Chin. Astron. Astrophys. 21/1:
1-18. Chinese version in Tianti Wuli Xuebao (Acta Astrophys.
Sinica), 1997, 17/3: 225-238. 

\bibitem{} Kiang T., 2001, Chin. Astron. Astrophys. 25/3:
141-146. 

\bibitem{} Kiang T., 2003, Chin. Astron. Astrophys. 27/3:
247-253.

\bibitem{} Longair M. S., 1984, Theoretical Concepts in
Physics, Oxford University Press, p.292.

\bibitem{} Misner C. W., Thorne K. S. and Wheeler J. A, 1973, 
Gravitation, Freeman, San Francisco, p.655.

\bibitem{} Rees M., 1967, MN, 135/4: 346.     

\bibitem{} Taylor E. F., and Wheeler J. A, 1963, Spacetime
Physics, Freeman, San Francisco, p.37.

\end{thebibliography}
\end{document}